\documentclass[fleqn,usenatbib]{mnras}

\usepackage{newtxtext,newtxmath}

\usepackage[T1]{fontenc}

\DeclareRobustCommand{\VAN}[3]{#2}
\let\VANthebibliography\thebibliography
\def\thebibliography{\DeclareRobustCommand{\VAN}[3]{##3}\VANthebibliography}

\usepackage{graphicx}	
\usepackage{amsmath}	
\usepackage{booktabs}
\usepackage{color,soul}

\title[On Bactrian glitch-size distributions]{On Bactrian glitch-size distributions}

\author[Anantharaman et al.]{S V Anantharaman,$^{1}$\thanks{E-mail: ananth.academic@gmail.com}
Dipankar Bhattacharya$^{1}$
\\
$^{1}$Department of Physics, Ashoka University, Sonipat, 131029, India\\
}
\date{Accepted XXX. Received YYY; in original form ZZZ}

\pubyear{\the\year{2024}}

\begin{document}
\label{firstpage}
\pagerange{\pageref{firstpage}--\pageref{lastpage}}
\maketitle

\begin{abstract}
A glitch is a rare and sudden increase in the otherwise steadily decreasing rotation rate of a pulsar. Its cause is widely attributed to the transfer of angular momentum to the crust of the star from the array of superfluid vortices enclosed within. The magnitude of such an increase defines the size of the glitch. The distribution of glitch sizes in individual pulsars, the power-law being the most argued for, is shrouded in uncertainty due to the small sample size. From a Bayesian perspective, we revisit the data for PSR J0537-6910, the pulsar with the most glitches, and find a bimodality in the distribution, reminiscent of the Bactrian camel. To understand this bimodality, we use a superfluid vortex simulator and study three independent neutron star paradigms: (i) Annular variation in pinning strength to account for the predicted differences between the crust and the core; (ii) Sectorial triggers to mimic local disturbances; and (iii) Stress-waves to model global disturbances. We find that annular variation in pinning introduces a bimodality in the glitch-size distribution and that sectorial triggers do so weakly. Stress-waves do not lead to any such features for the range of parameters tested. This provides us with new insights into the effects of various perturbations on the vortex dynamics and the glitch statistics of neutron stars.
\end{abstract}

\begin{keywords}
stars: neutron -- stars: interiors -- stars: rotation -- pulsars: general -- pulsars: individual: PSR J0537-6910
\end{keywords}



\section{Introduction}

Pulsars exhibit great diversity in their glitching behaviour. The associated statistics are no different. The progression of a glitch, the subsequent relaxation, the resulting glitch size, and the associated waiting times span a wide variety between different pulsars (\cite{antonopoulou2022}). The glitch data for individual pulsars is also scarce. The most glitching pulsar, PSR J0537-6910, has exhibited 53 such events over 20 years. The most frequently monitored ones, PSR J0534+2200 (Crab), and PSR J0835-4510 (Vela), have had 30 and 25 glitches, respectively. Most others have had only 10 glitches or fewer (\cite{basu2022}). Hence, studies of glitch statistics have concentrated on data collected from all pulsars, with fewer works concerned with individual statistics (\cite{eya2019,konar2014,fuentes2017}). Power-law, log-normal, Gaussian, and mixed glitch-size distributions have all been considered as fits for data corresponding to different pulsars (\cite{fuentes2019,espinoza2014,howitt2018}). A unified description to explain the host of observational data has been a long-standing challenge in pulsar physics. Crucial to such a synthesis is understanding the internal dynamics of pulsars.

Matter in the interior of pulsars (rotating neutron stars) is expected to be in a superfluid state which supports rotation by the formation of quantised vortices. In the crust of the star, such vortices are held in place by nuclear lattice sites or defects in the lattice. Glitches are thought to be caused primarily by the collective release of several vortices from their respective traps. The power-law distribution of glitch sizes has been the primary prediction of a variety of theoretical models including Gross-Pitaevskii simulations and cellular automaton approximations (\cite{warszawski2008,warszawski2011}). However, a recent work, in which glitches were simulated using a hydrodynamic two-dimensional N-body superfluid vortex setup, suggests that both log-normal and power-law distributions are valid considerations in the regime of large glitches (\cite{howitt2020}). This disconnect between the theoretical and the observational glitch statistics prompts us to take a three-pronged perspective that explores and eventually unites observation, theory, and computational modelling.

In Section 2, we analyse the glitch-size data of PSR J0537-6910. In Section 3, we briefly introduce the simulator and study glitches resulting from vortex avalanches triggered solely by the Magnus force, the standard case. By leveraging the idea of generative models, we also strike a connection between the fundamental processes inside the star to the resulting size distribution of glitches. In Section 4, motivated by the results of Section 2, we introduce into the simulator three new considerations: annular pinning, sectorial triggers, and stress-waves, each of which mimics the effects of a subset of phenomena expected to occur in neutron stars. We study the change in the distribution in each case and elucidate the insights thus derived. Further discussion and a summary of our work are provided in Section 5.

\section{Revisiting observations}

\begin{figure}
  \centering
  \includegraphics[width=\linewidth]{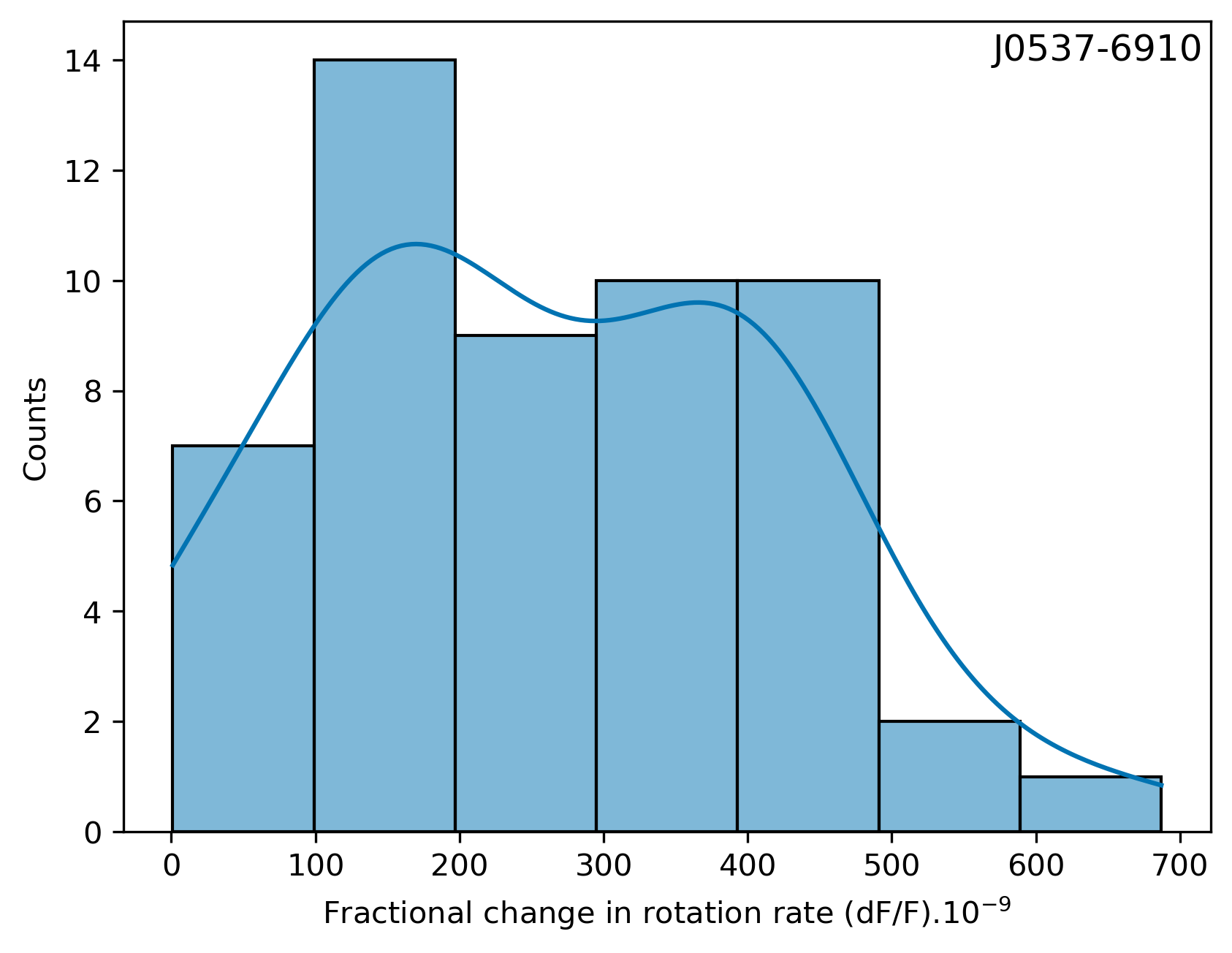}
  \caption{The observed distribution of glitch sizes for PSR J0547-6910. The solid line represents the corresponding Kernel Density Estimate.}
  \label{fig:observation}
\end{figure}

The glitch-size distribution of PSR J0537-6910, the pulsar with the most glitches, along with its Kernel Density Estimate (KDE) (\cite{parzen1962}) is presented in Fig.~\ref{fig:observation}. A bimodality is apparent in the KDE corresponding to PSR J0537-6910. We test this hypothesis using two methods.

\subsection{Bayesian Information Criterion (BIC)}

In this section, we compare the BIC values obtained for various bimodal distributions with those obtained for other standard fits.

The essential component of the BIC is the likelihood function, calculated for a model and data set pair, with a penalty imposed for each free parameter in the model (\cite{schwarz1978}).

\begin{equation}
~~~~~~~~~~~~~~~~~~~~~~~~~~~~~~~~~~~~
\mathrm{BIC}=-2 \ln (\widehat{L})+k \ln (n)
\label{bic}
\end{equation}
where $\widehat{L}$ is the maximized value of the likelihood function for the chosen model, $n$ is the number of observations available and $k$ is the number of parameters estimated by the model. For a given set of observations, if two models differ in their BIC by a value of 10, then the model with the lower BIC is about 150 times as likely as the other, thus making the model selection reliable. On the other hand, if the BIC differs by 5 or less, the selection of models is unreliable (\cite{kass1995}). The model with the lowest of the BIC values is generally preferred.

Using this measure, we test various models against the glitch-size data for PSR J0537-6910. We employ the LMFIT package available for python to achieve the same (\cite{newville2015}). We consider three bimodal distributions: double log-normal, double Gaussian, and lognormal-Gaussian mixture. The forms of the bimodal distributions that we have used are obtained by a straightforward addition of the well-known standard distributions (\cite{hays1971}). For instance, the double log-normal is obtained by adding two log-normal distributions. The bimodal distributions are then compared with a few standard distributions of interest: Gaussian, log-normal, exponential, and power-law. The distribution functions and the corresponding BIC values are mentioned in Table~\ref{tab:observation_bic}. The best fit parameters are not displayed in the table as they are not of consequence to the present discussion.

\begin{table}
    \centering
    \caption{The BIC corresponding to various models, tested against the glitch-size data of PSR J0537-6910.}
    \label{tab:observation_bic}
    \begin{tabular}{@{}ll@{}}
    \toprule
    \textbf{Distribution} & \textbf{BIC} \\ \midrule
    Lognormal-Gaussian    & -27.65       \\
    Double log-normal    & -23.12       \\
    Double Gaussian      & -3.408        \\
    Gaussian              & -0.266        \\
    Log-normal             & 4.612        \\
    Exponential           & 8.387        \\
    Power-law             & 11.88        \\ \bottomrule
    \end{tabular}
\end{table}

\begin{figure}
  \centering
  \includegraphics[width=\linewidth]{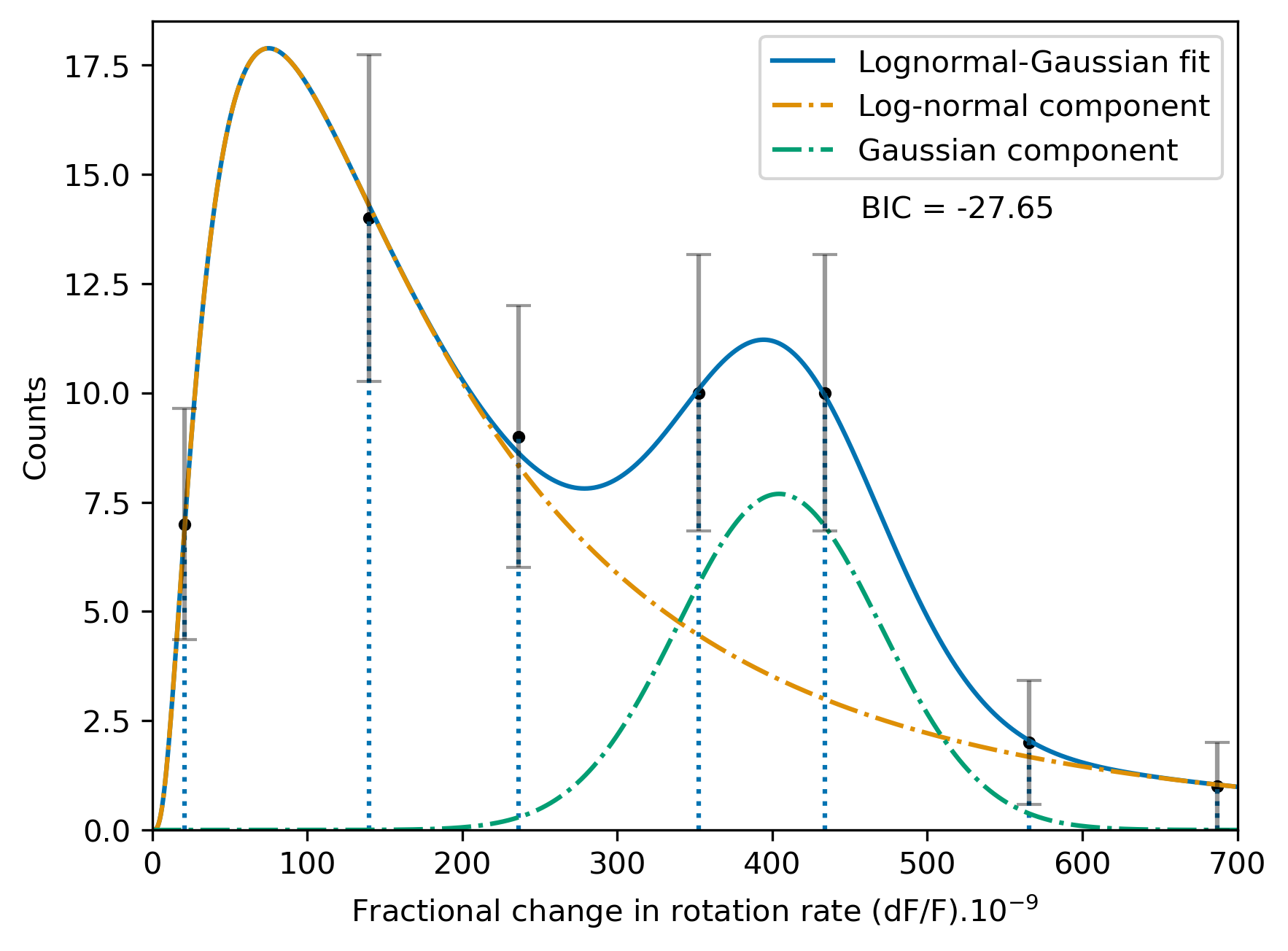}
  \caption{Fitting of the distribution with the lowest BIC to the glitch-size data of PSR J0537-6910. The solid line is the lognormal-Gaussian mixture with best fit parameters. The two unimodal components of the distribution are shown as dashed curves. The dotted vertical lines represent the bins of the histogram. The positions of the lines on the horizontal axis are decided by the means of the entries in the respective bins. The heights are simply the corresponding counts. The errors in counts are approximated as $\sqrt{n}$, where n is the number of observations in the respective bins.}
  \label{fig:observation_bimodality}
\end{figure}

We find that the bimodal distributions each have a BIC much lower than those of the standard distributions. Among the bimodal models, the lognormal-Gaussian mixture has the lowest BIC, separated from the double log-normal by a value of 4.20, and from the double Gaussian by a value of 12. This indicates that the glitch-size distribution for the pulsar considered is best described by a lognormal-Gaussian or a double log-normal. Fig.~\ref{fig:observation_bimodality} shows the best-fit distribution for the data. 

\subsection{Hartigan's Dip Test (HDT)}

HDT is a statistical tool to quantify multimodality in a given dataset (\cite{hartigan1985_1};\cite{hartigan1985_2}). The test produces two measures: (i) the dip statistic; and (ii) the p-value. Both of these measures range from zero to one. The dip statistic captures the prominence of the dip in the distribution, an essential indicator of multimodality. The p-value conveys the associated confidence level. A lower p-value indicates a higher probability of the distribution being multimodal.

Performing the dip test on the glitch-size data for PSR J0537-6910 lends a dip statistic of 0.053, and a p-value of 0.313. This implies that the distribution under study is roughly 70 percent likely to be bimodal. Hartigan's measure is clearly more conservative than the Bayesian approach used in the previous subsection.

The results of the above methods are in contrast to that presented in \cite{howitt2018}, where no convincing evidence for multimodality was found.

\section{The simulator}

Simulations of superfluid vortex arrays in neutron stars come in two varieties. The first of them takes the perspective that the star is composed of a few vortices ($\sim 10^{3}$) (\cite{warszawski2011,howitt2020}). The second assumes that the star is composed of a large number of vortices and attempts to model only a small fraction of the superfluid present within the star (\cite{liu2024}). Since our interest lies in understanding the collective dynamics of vortices and their impact on the rotation of the star, we prefer the former approach. Here, we adopt a computationally simple hydrodynamic prescription to simulate superfluid vortices in two dimensions, as detailed in \cite{howitt2020}. The units for all physical quantities mentioned below are the same as those in the provided reference.

\subsection{Construction}

A star (container) of radius $R$ with a background lattice of pinning sites and almost uniformly distributed vortices sets the stage where the dynamics unfolds. The interaction between the vortex and the bulk superfluid is endowed with a small but finite dissipation, and the star experiences constant deceleration. The positions of the vortices, then, evolve according to equation~(\ref{eq:eom}).

\begin{equation}
    ~~~~~~~~~~~~~~~~~~~~~~~~~~~~~~~~~
    \frac{\mathrm{d}}{\mathrm{d} t}\left(\begin{array}{l}
x_{i} \\
y_{i}
\end{array}\right)=\mathcal{R}_{\phi}\left(\begin{array}{c}
v_{i, x} \\
v_{i, y}
\end{array}\right)
\label{eq:eom}
\end{equation}
where
\begin{equation}
    v_{i, x}=-\sum_{j \neq i} \frac{\kappa y_{i j}}{r_{i j}^{2}}+\sum_{j=1}^{N} \frac{\kappa y_{i j, \text { image }}}{r_{i j, \text { image }}^{2}}+\Omega_c y_{i}-\sum_{pin} \frac{\partial V\left(\mathbf{x}_{i}-\mathbf{x}_{pin}\right)}{\partial y_{i}}
    \label{eq:vx}
\end{equation}

\begin{equation}
    v_{i, y}=\sum_{j \neq i} \frac{\kappa x_{i j}}{r_{i j}^{2}}-\sum_{j=1}^{N} \frac{\kappa x_{i j, \text { image }}}{r_{i j, \text { image }}^{2}}-\Omega_c x_{i}+\sum_{pin} \frac{\partial V\left(\mathbf{x}_{i}-\mathbf{x}_{pin}\right)}{\partial x_{i}} .
    \label{eq:vy}
\end{equation}

Here, $\mathbf{x}_{i}$ is the position vector of the vortex, whose components are $x_{i}$ and $y_{i}$, calculated with respect to the origin at the centre of the container. The terms on the right hand side in equations~(\ref{eq:vx},\ref{eq:vy}) account for the effect of other vortices, the presence of a boundary, the rotation of the star, and the influence of the pinning sites, in that order, on the vortex under consideration. The rotation matrix, $\mathcal{R}_{\phi}$, accounts for the dissipation in the system.

The angular speed of the superfluid is related to the distribution of the vortices through equation~(\ref{eq:superfluidrate}) and the evolution of the rotation rate of the container is described by equation~(\ref{eq:crustrate}).
\begin{equation}
~~~~~~~~~~~~~~~~~~~~~~~~~~~~~~~~~~~~~
\Omega_s=\frac{k}{I_s} \sum_{i=1}^N\left(R^2-r_i^2\right)
\label{eq:superfluidrate}
\end{equation}
where $k$ is a constant fixed by the vortex positions and the rotation rate at the start of the simulation.
\begin{equation}
~~~~~~~~~~~~~~~~~~~~~~~~~~~~~~~~~~~~~
\frac{d \Omega_c}{d t}=N_{\mathrm{ext}}-I_{\mathrm{rel}} \frac{d \Omega_s}{d t}
\label{eq:crustrate}
\end{equation}
For a detailed note on the equations and notation, refer to \cite{howitt2020}. Table~\ref{tab:parameters} details the parameters used, together with those relevant to the simulations described in Section 4. Henceforth, we refer to the construction presented in this section as the standard case. 

\begin{table*}
  \caption{Summary of the parameters used in the standard, annular pinning, sectorial trigger and stress-wave simulations. Top panel: Parameters used in the standard simulations and common to all others, except for the pinning strength, which is different for the annular pinning model. Bottom panel: Additional variation-specific parameters.}
  \label{tab:parameters}
  \begin{tabular}{@{}lll@{}}
  \toprule
  \textbf{Standard parameters} & \textbf{} &  \\ \midrule
  \begin{tabular}[c]{@{}l@{}}Radius of  the star\\ $R = 10$\\ \\ Number of  vortices \\ $N_v = 2000$\\ \\ Dissipation strength\\ $\phi = 0.1$\\ \\ Ratio of superfluid/crust\\ moments of inertia\\ $\mathrm{I_{rel}} = 1$\\ ~\end{tabular} & \begin{tabular}[c]{@{}l@{}}Number of pinning sites\\ $N_{\mathrm{pin}} = 20000$\\ \\ Pinning strength\\ $V_0 = 2000$\\ \\ Influence of pinning site\\ $\xi = 0.001~R$\\ \\ Runtime of simulation\\ $T_{\mathrm{run}} = 2000~ T_0$\\ ~\\ ~\end{tabular} & \begin{tabular}[c]{@{}l@{}}Initial rotation rate\\ $\Omega_0 = N/R^{2} = 20$\\ \\ Initial rotation period\\ $T_0 = 2\pi/\Omega_0 = 0.314$\\ \\ Spin-down rate\\ $\mathrm{N_{ext}} = -2.5 \times 10^{-4} ~\Omega_0/T_0$\\ \\ Default integration timestep\\ $dt = 0.005 ~T_0$\\ ~\\ ~\end{tabular} \\ \midrule
  \textbf{Annular pinning} & \textbf{Sectorial trigger} & \textbf{Stress-wave} \\ \midrule
  \begin{tabular}[c]{@{}l@{}}Radius of inner disc\\ $r_{\mathrm{annulus}}=  7.07$\\ \\ Pinning strength for:\\ inner disc $V_1$\\ outer annulus $V_2$\\ \\ \\ \\ We run sets of simulations with\\ $V_1, ~V_2 $ as:\\ 1000, ~2000\\ 2000, ~1000\\ 3000, ~2000\\ 2000, ~3000\\ ~\end{tabular} & \begin{tabular}[c]{@{}l@{}}Number of triggers\\ $n_{\mathrm{trig}}$\\ \\ Duration of triggers\\ $T_{\mathrm{trig}} = 1~T_0$\\ \\ Spatial extent of triggers\\ $\mathrm{\textit{region}}$\\ \\ We run sets of simulations with\\ $n_{\mathrm{trig}}, ~\mathrm{\textit{region}}$ as:\\ 40, ~full sector\\ 60, ~outer sector\\ ~\\ ~\\ ~\end{tabular} & \begin{tabular}[c]{@{}l@{}}Number of triggers\\ $n_{\mathrm{trig}}$\\ \\ Duration of triggers\\ $T_{\mathrm{trig}} = 1~T_0$\\ \\ Pinning strength reduction factor\\ $\gamma$\\ \\ We run sets of simulations with\\$n_{\mathrm{trig}}, \gamma$ as:\\40, 0.7\\20, 0.5 \\~ \\~ \\~ \\ \end{tabular} \\ \bottomrule
  \end{tabular}
\end{table*}
The simulations are initiated with all the vortices being pinned, and ensuring that their distribution is such that the rotation of the superfluid is in equilibrium with the rotation of the star. The spindown of the star and the pinning of the vortices lead to a gradually increasing Magnus force on the vortices. The stress caused by the spindown is captured by the sum of the first three terms in equations~(\ref{eq:vx}, \ref{eq:vy}). When this exceeds the pinning stress described by the last term, the vortices unpin, defined as moving out of the characteristic radius of a pinning site. Under the right local conditions, an avalanche occurs. The redistribution of the vortices, with a radially outward bias due to the Magnus force, causes a decrease in the rate of rotation of the superfluid and a corresponding increase in that of the container, according to equations (\ref{eq:superfluidrate}, \ref{eq:crustrate}). We record the positions of the vortices and the angular speed of the container ($\Omega_{c}$) at every time step. 

Animation A1 shows the evolution of the vortex array during a typical glitch and the associated change in the rate of rotation of the container. In the animation, the yellow dots are all the vortices that participate in the glitch. The grey dots represent the vortices that remain pinned. The white tails represent movement, and their lengths convey the instantaneous speed of the vortices. At the beginning of the glitch, most vortices move radially outward. This leads to a void in the region which, in turn, changes the local stresses. The result is a radially inward and partly azhimuthal collective motion, causing the vortices to fill the void almost uniformly and marking the end of the glitch. This behaviour has also been observed in the Gross-Pitaevskii simulations recently reported by \cite{liu2024}. All animations referred to in this article are available for viewing online on the publisher's website.

\subsection{Simulational glitch statistics}

The detection of glitches is performed by tracking the slope of the instantaneous rotation rate of the container. A glitch is defined to begin when the slope changes from negative to positive and end when the slope changes back to being negative. To faithfully reflect the statistics of the system, the detection mechanism considers glitches of all sizes with no minimum threshold.

\begin{figure}
  \centering
  \includegraphics[width=\linewidth]{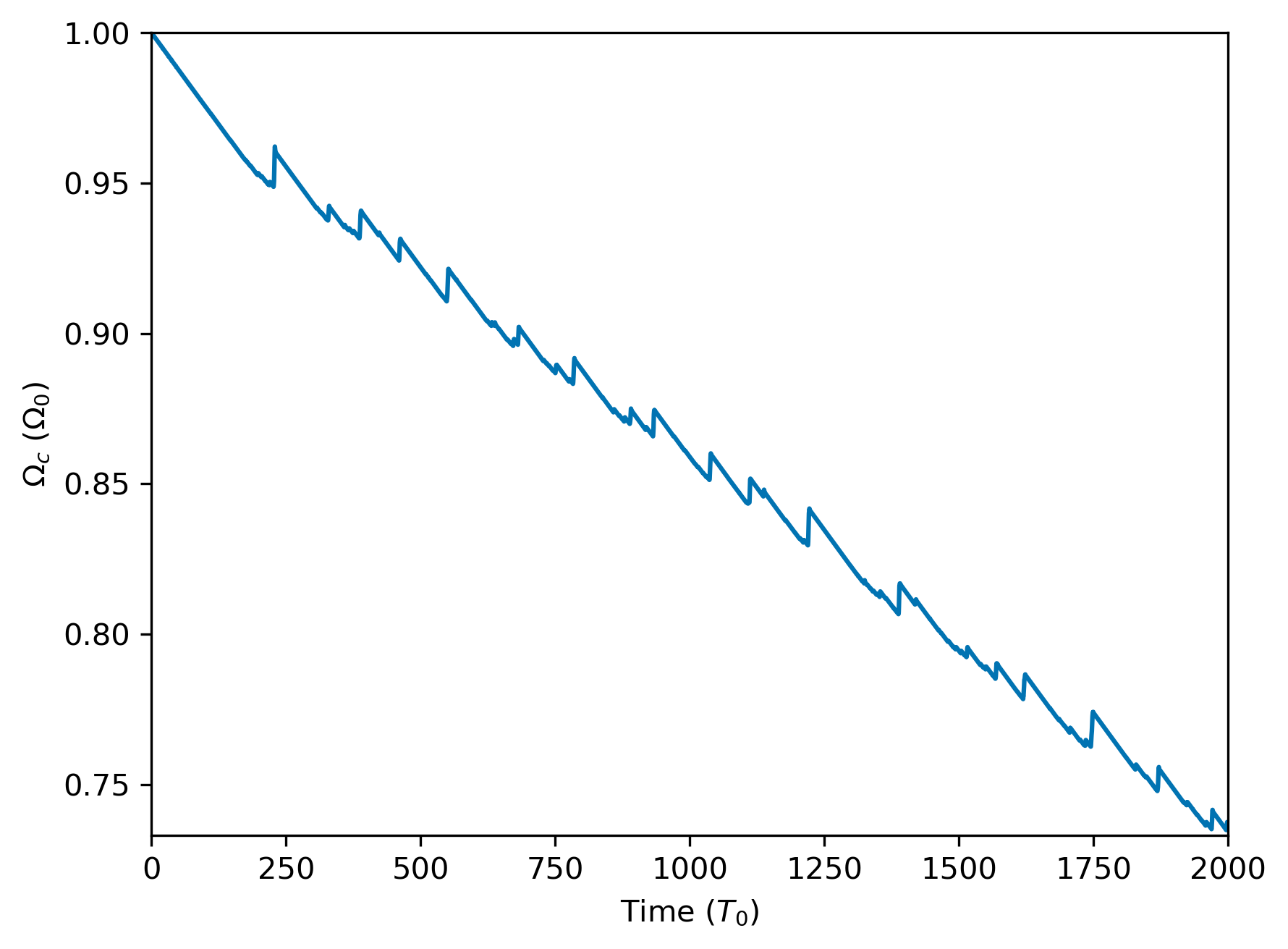}
  \includegraphics[width=\linewidth]{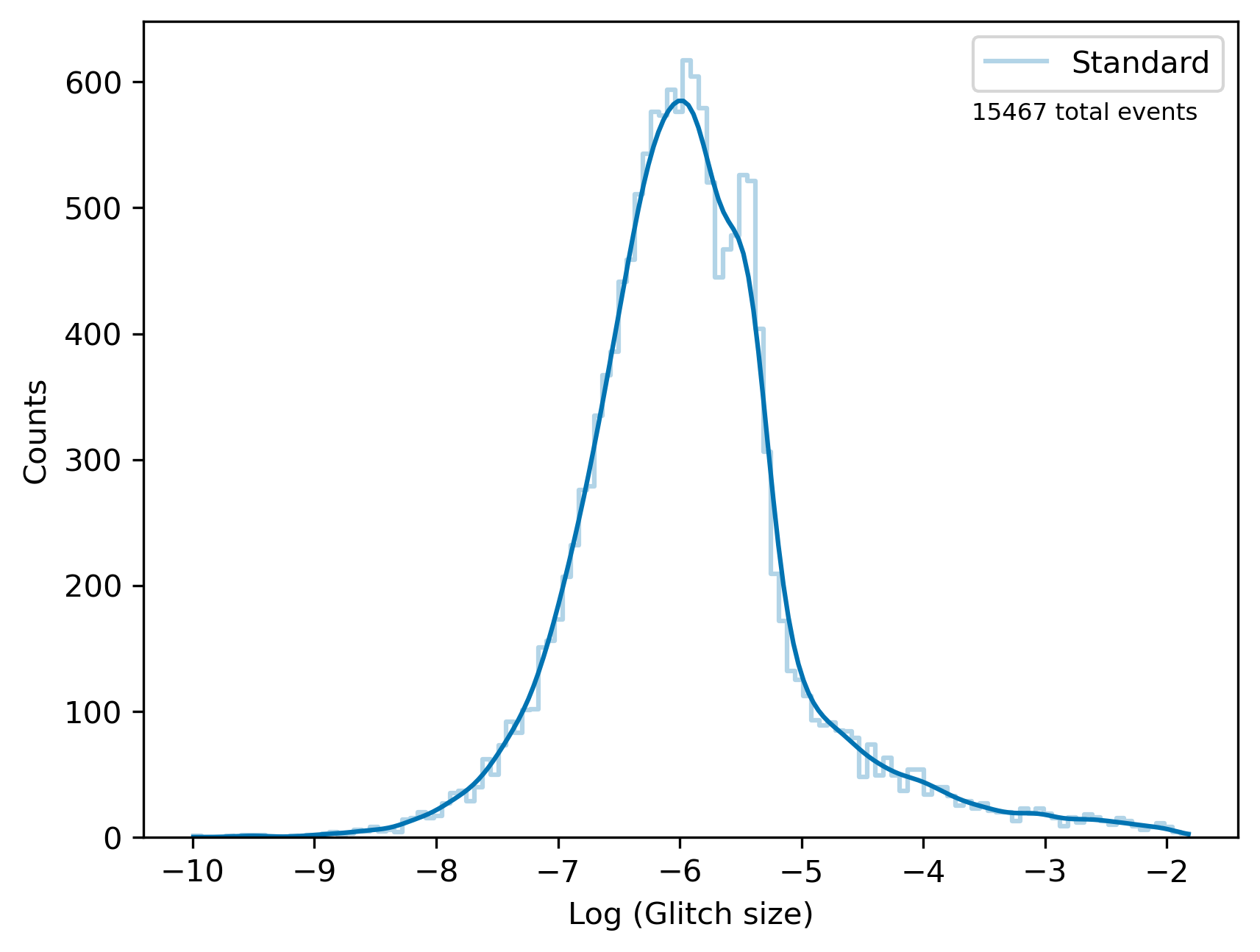}
  \includegraphics[width=\linewidth]{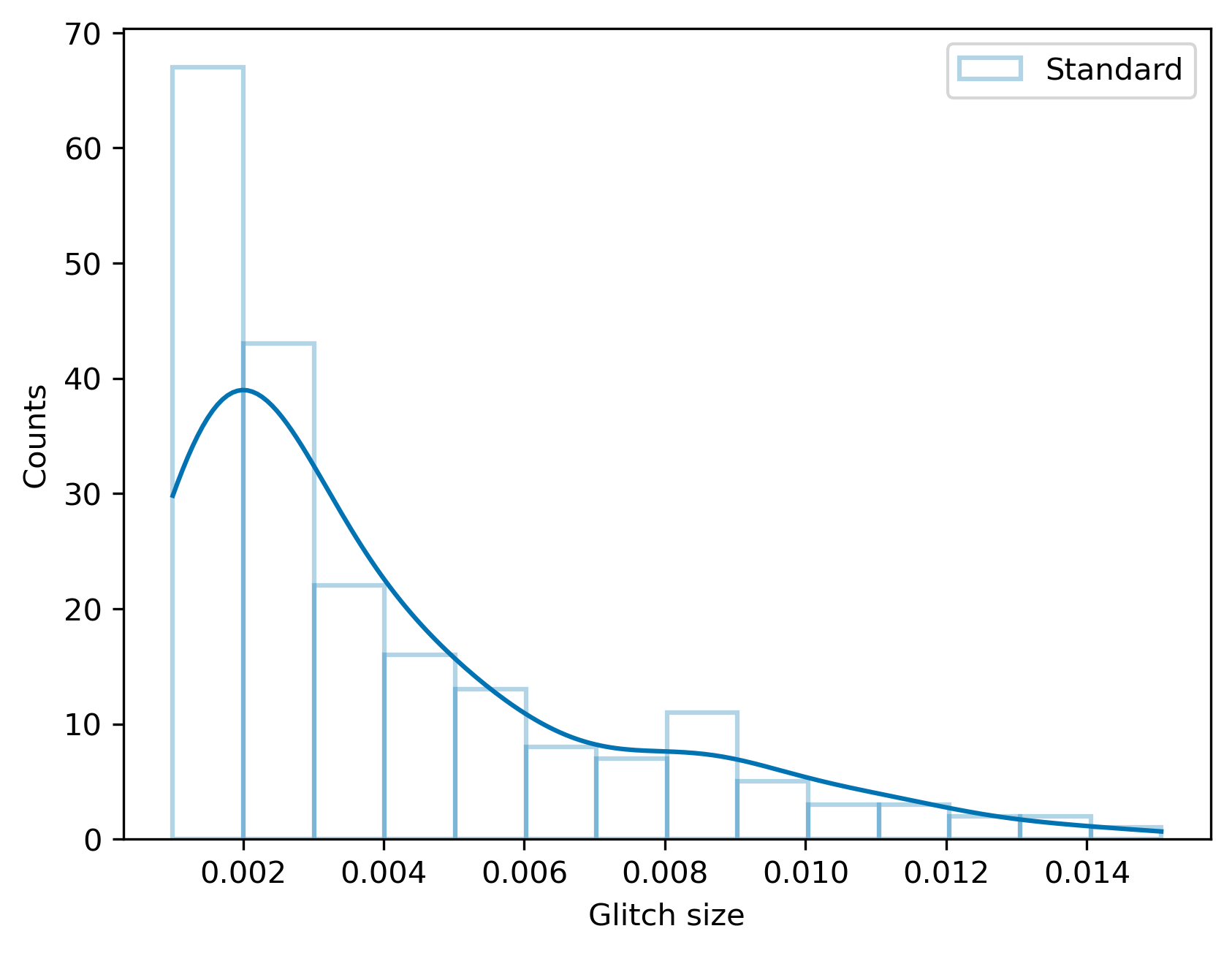}
  \caption{Data from standard simulations. Top panel: Evolution of the star's rate of rotation in one of the runs. Several glitches can been seen. Middle panel: The distribution of all the glitches recorded by the detection scheme, with no cutoff at the lower end. Bottom panel: Distribution of only the large glitches.}
  \label{fig:standard}
\end{figure}

Fig.~\ref{fig:standard} shows the slew of glitches produced in a single run of the standard simulation, the glitch-size distribution resulting from five such runs, and the distribution of only the large glitches. We define the size of a glitch to be $\Delta\Omega_{c}/\Omega_{0}$. In a large glitch, several vortices collectively migrate over distances much greater than the characteristic pinning radius. In the standard paradigm, such events involve tens of vortices being displaced across a significant fraction of the star's radius, resulting in glitches having sizes greater than $10^{-3}$. Throughout this article, we obtain reliable simulational statistics by collating the data generated by five runs of the corresponding set-up with only the initial positional configuration of the vortex array changing between the runs. For the standard case, this totals about 15000 glitches forming the full distribution, with approximately 200 of them constituting the distribution of large glitches.

Akin to the procedure followed in Section 2, we test various statistical models against the glitch-size data obtained from the simulations. The corresponding BIC values are reported in Table~\ref{tab:sim_standard_bic}. They indicate that the distribution is one where the body is described best by a log-normal. The tail at the higher end is also explained well by a log-normal, and, to a lesser degree of confidence, by an exponential or a power-law.

In the simulations, we find that none of the vortices is unpinned for glitches of a size smaller than approximately $10^{-5}$. To understand this, we place bounds on the transfer of angular momentum resulting from a few simple processes. From equations (\ref{eq:superfluidrate}, \ref{eq:crustrate}), we obtain:

\begin{equation}
~~~~~~~~~~~~~~~~~~~~~~~~~~~~~~~~~~~~~
\Delta\Omega_s=\frac{k}{I_s} \sum_{i=1}^N\left(-2r_i\right) \Delta r_{i}
\label{eq:superfluidrate_diff}
\end{equation}

\begin{equation}
~~~~~~~~~~~~~~~~~~~~~~~~~~~~~~~~~~~~~
\Delta\Omega_c=N_{\mathrm{ext}}\Delta t-I_{\mathrm{rel}} \Delta\Omega_s
\label{eq:crustrate_diff}
\end{equation}

where $\Delta$ represents a small change in the associated quantity.

The spin-down decreases $\Omega_{c}$ at the end of every time step ($dt$). For a glitch to be detected, $\Delta\Omega_{c}$ should be positive. Equation \ref{eq:crustrate_diff} sets the minimum difference required in the superfluid rotation rate ($\Delta\Omega_{\mathrm{s,min}}$) between successive instants in time for a glitch to occur. For the parameters of our simulations, $\Delta\Omega_{\mathrm{s,min}} = N_{\mathrm{ext}}(dt)/I_{\mathrm{rel}} = - 1.25 \times 10^{-6} \Omega_{0}$. 

A single vortex that hops one pinning site, radially outward, near the centre of the star, and one that hops near the boundary of the star would lead to a decrease in $\Omega_{s}$ roughly by $10^{-7}~\Omega_{0}$ and $10^{-5}~\Omega_{0}$, respectively. In the former case, no glitch would be recorded. In the latter case, a glitch would be detected, having a size of approximately $10^{-5}~\Omega_{0}$, imposing a lower bound on the size of a glitch resulting from a collective motion of vortices. Thus, we find that the body of the standard distribution is composed of micro-glitches caused by the small changes in the locations of the vortices at every instant in time, which in turn is driven by the continuously decreasing Magnus force. The tail of the distribution at the higher end ($\sim 10^{-3}$) corresponds to the collective motion of vortices. It is this limit that we shall probe in the following sections.

\begin{table}
\centering
\caption{The BIC corresponding to various models,
tested against the glitch-size data obtained from standard simulations.}
\label{tab:sim_standard_bic}
\begin{tabular}{@{}lll@{}}
\toprule
\textbf{Data set} & \textbf{Distribution} &  \textbf{BIC} \\ \midrule
Full & Log-normal    & 284.7       \\
Full & Gaussian      & 384.5       \\
Full & Exponential    & 491.1       \\
Large & Log-normal              & -7.761        \\
Large & Exponential           & 1.247        \\
Large & Power-law             & 1.444        \\ \bottomrule
\end{tabular}
\end{table}

The connection between the simulator and the resultant glitch statistics is discussed in Appendix A.

\section{Variations}

The observed glitch sizes of PSR J0537-6910 are best represented by a bimodal distribution, either the lognormal-Gaussian mixture or the double log-normal. On the other hand, the simulator produces a unimodal distribution, best approximated by a log-normal, irrespective of whether the entire dataset or only the large glitches are considered. The bimodality in observation could indicate the existence of two characteristic energy scales. Three aspects of pulsars that could possibly introduce such effects are: (i) the differences in the properties of the crust and the core; (ii) disturbances in the star that unpin a large number of vortices in a localised region; and (iii) disturbances that unpin the most stressed vortices across the star. The involvement of the core superfluid and trigger mechanisms have previously been suggested as explanations for large glitches (\cite{andersson2012}).

In subsections 4.1 to 4.3, we incorporate these considerations into our simulator and present the resulting glitch-size distributions. In subsection 4.4, we quantify the bimodality observed in each case and comment on the same.

\subsection{Annular pinning}
Pinning of vortices in the crust and the core of neutron stars is predicted to be provided by nuclear centres and magnetic flux tubes, respectively. The difference in the number density of pinning sites lends different values of lag for which vortices get unpinned in the two regions. This critical lag is estimated to be $10$ rad/s for the crust and $10^{-2}$ rad/s for the core (\cite{sauls1989}). The situation is further complicated by the flux-tubes and the vortices possibly tangling and twisting around each other (\cite{srinivasan1990}). This would lead to a significantly larger critical lag for the core than predicted in \cite{sauls1989}. The exact value is not yet available because the picture is far from clear. So, we take the two values of the critical lag to be approximately of the same order of magnitude.

\begin{figure}
  \centering
  \includegraphics[width=0.7\linewidth]{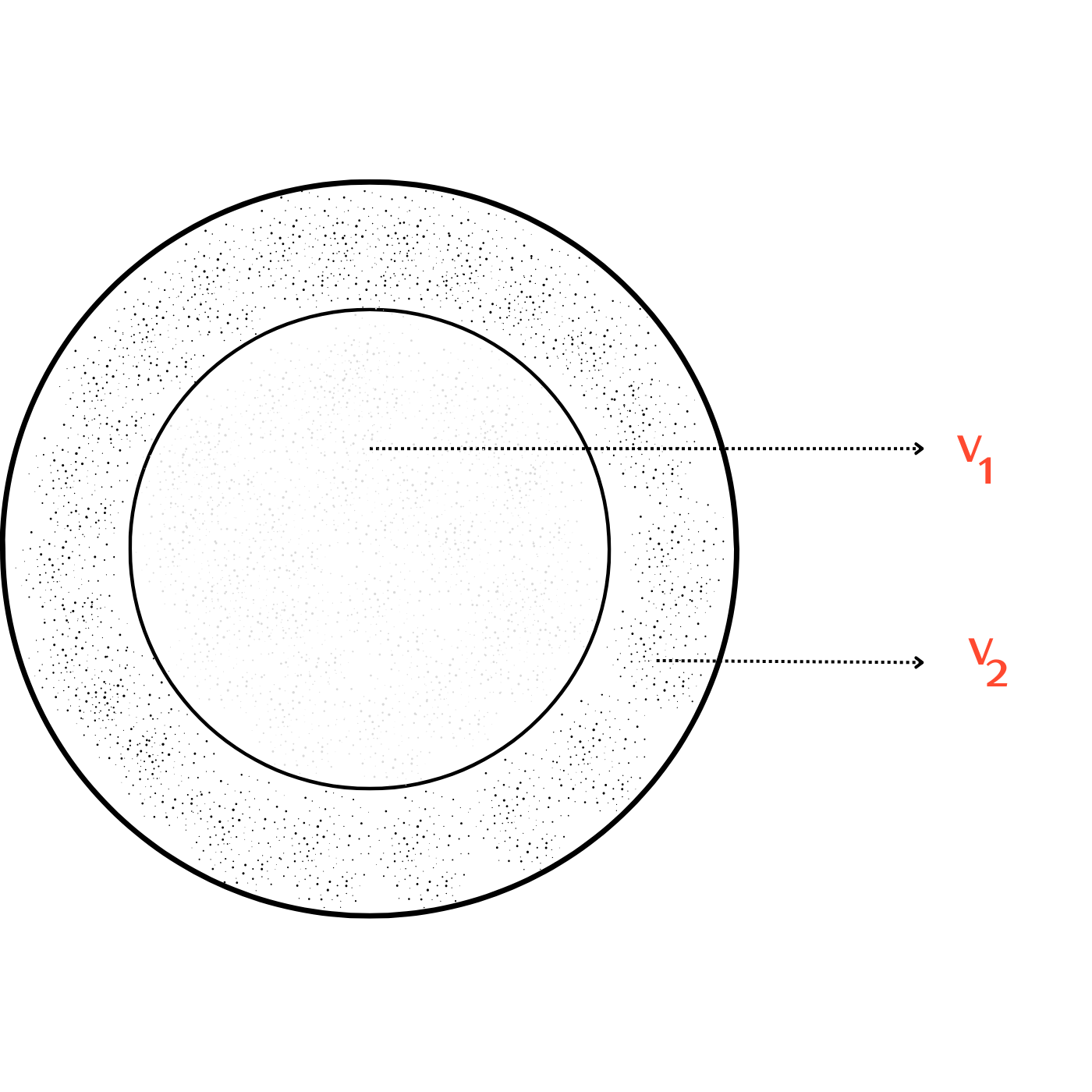}
  \caption{A schematic of annular variation in pinning.}
  \label{fig:annular_schematic}
\end{figure}

To capture the physics of such a system, we simulate a model with an annular variation of the pinning strength. We divide the star into two regions, an inner disc and an outer annulus, with pinning sites having strengths $V_{1}$ and $V_{2}$ respectively. See Fig.~\ref{fig:annular_schematic}. The standard scenario discussed in Section 3 would be reproduced if the pinning strengths in both the regions were set to the same value of 2000.

Here, we choose to explore four combinations of pinning strengths ($V_1$, $V_2$): (1000, 2000); (2000, 1000); (3000, 2000); and (2000, 3000). In all these simulations, the areas of the inner disc and that of the outer annulus are the same. Thus, the population of the two species of pinning sites is in a ratio of 1:1.

\begin{figure}
  \includegraphics[width=\linewidth]{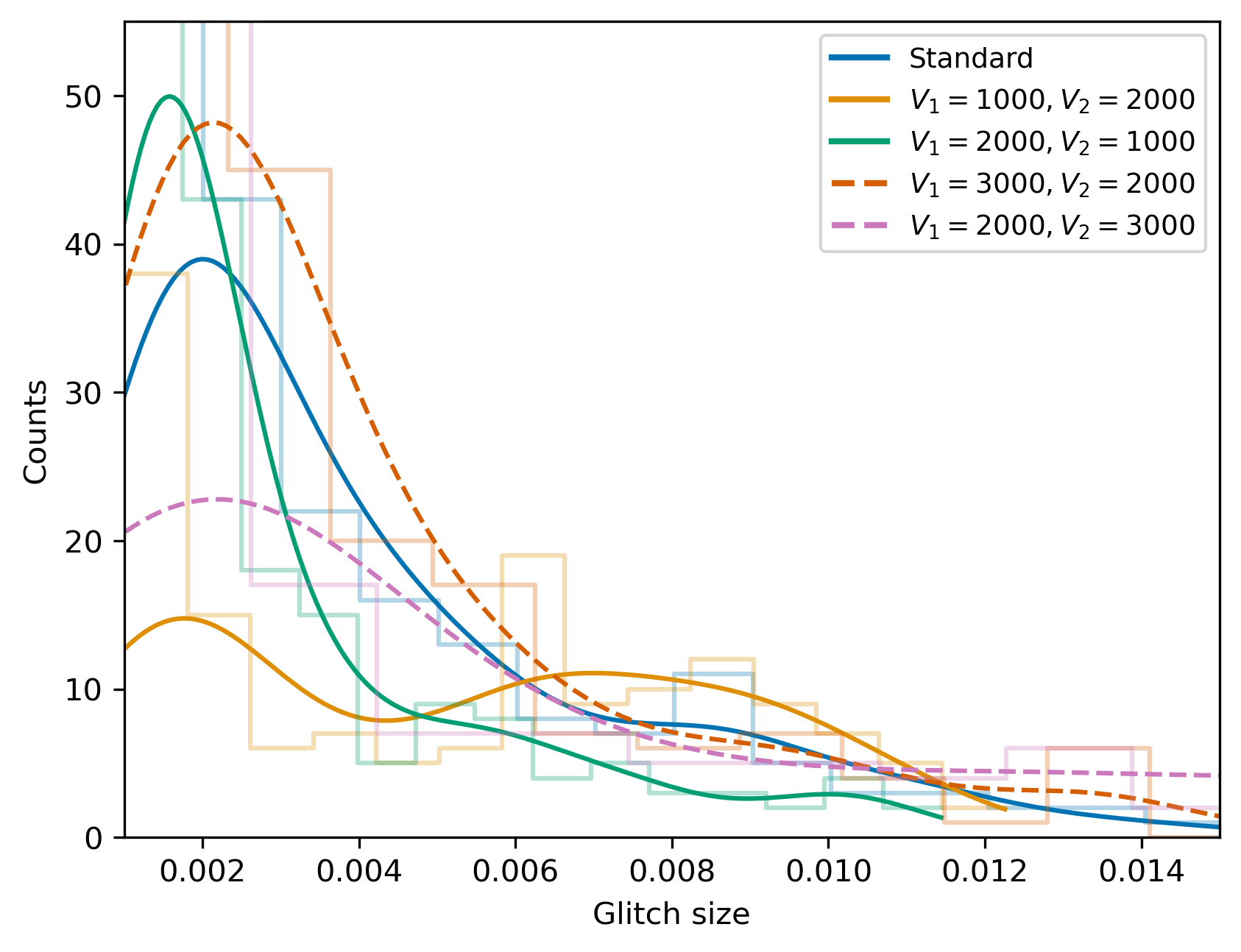}
  \caption{Distributions of large glitches resulting from the annular pinning simulations compared with the standard one. The solid lines and the dashed lines are the corresponding Kernel Density Estimates. The relevant pinning strengths are mentioned within the plot.}
  \label{fig:annular_distributions}
\end{figure}

In Fig.~\ref{fig:annular_distributions}, we juxtapose the various size distributions resulting from the annular pinning schemes. The associated statistics are presented in Table~\ref{tab:simulator_statistics}. The size of the largest glitch, that of the smallest glitch, and the mean of the entire distribution are calculated directly from the raw data. The mode of the distribution is studied on the logarithmic scale, reflecting where the peaks occur in Fig.~\ref{fig:annular_distributions}.

\begin{table}
\centering
\caption{Statistics of large glitches resulting from the standard and the annular pinning simulations. For reference, the statistics for two other pertinent cases are also provided.}
\label{tab:simulator_statistics}
\begin{tabular}{@{}lll@{}}
\toprule
\textbf{Configuration ($V_1$, $V_2$)} & \textbf{Max ($10^{-2}$)} & \textbf{Mean ($10^{-3}$)}\\ \midrule
Standard             & 1.505                 & 3.948\\
1000, 2000           & 1.226                 & 5.220\\
2000, 1000           & 1.144                 & 4.072\\
3000, 2000           & 1.934                 & 2.867\\
2000, 3000           & 2.354                 & 5.991\\
1000, 1000           & 0.609                 & 2.376\\ 
3000, 3000           & 2.615                 & 5.027\\ 
\bottomrule
\end{tabular}
\end{table}

A change in the pinning strength of one of the pinning regions does affect the size of the largest glitch, albeit in a marginal manner. Compared to the standard case, the maximum glitch size is smaller for simulations with one of the pinning strengths having a value $V=1000$ and larger for those with $V=3000$.

The mean of the distribution, on the other hand, does not follow the simple positive correlation noted in \cite{howitt2020}. That is, the presence of pinning sites with a smaller pinning strength does not simply translate to a reduction in the mean glitch size. The positive correlation holds only when the pinning strength is changed uniformly throughout the star.

\subsection{Sectorial triggers}

There is no consensus about the exact nature of the trigger that causes a glitch. The threshold-triggered avalanche mechanism described earlier is certainly a possibility. However, the diverse observational features exhibited by glitches in the same pulsar have led to the suggestion that several triggers could be acting in the same neutron star. For instance, a number of instabilities can be present in the superfluid, which could lead to both classical and quantum turbulence. A change between streamlined and turbulent flow could also trigger a glitch (\cite{glampedakis2008,khomenko2019,peralta2006}). Crustquakes could also be considered in the same vein, the initial models of which proposed that stress-related crust breaking was the sole reason for the star's spin up (\cite{ruderman1969}). Although these models did not explain the glitch data corresponding to several pulsars, the effect of such quakes on the vortex array is of interest and is not well understood (\cite{haskell2015,rencoret2021}). To model such events which lead to mass unpinning in a portion of the star, we introduce sectorial triggers.

\begin{figure}
  \centering
  \includegraphics[width=0.7\linewidth]{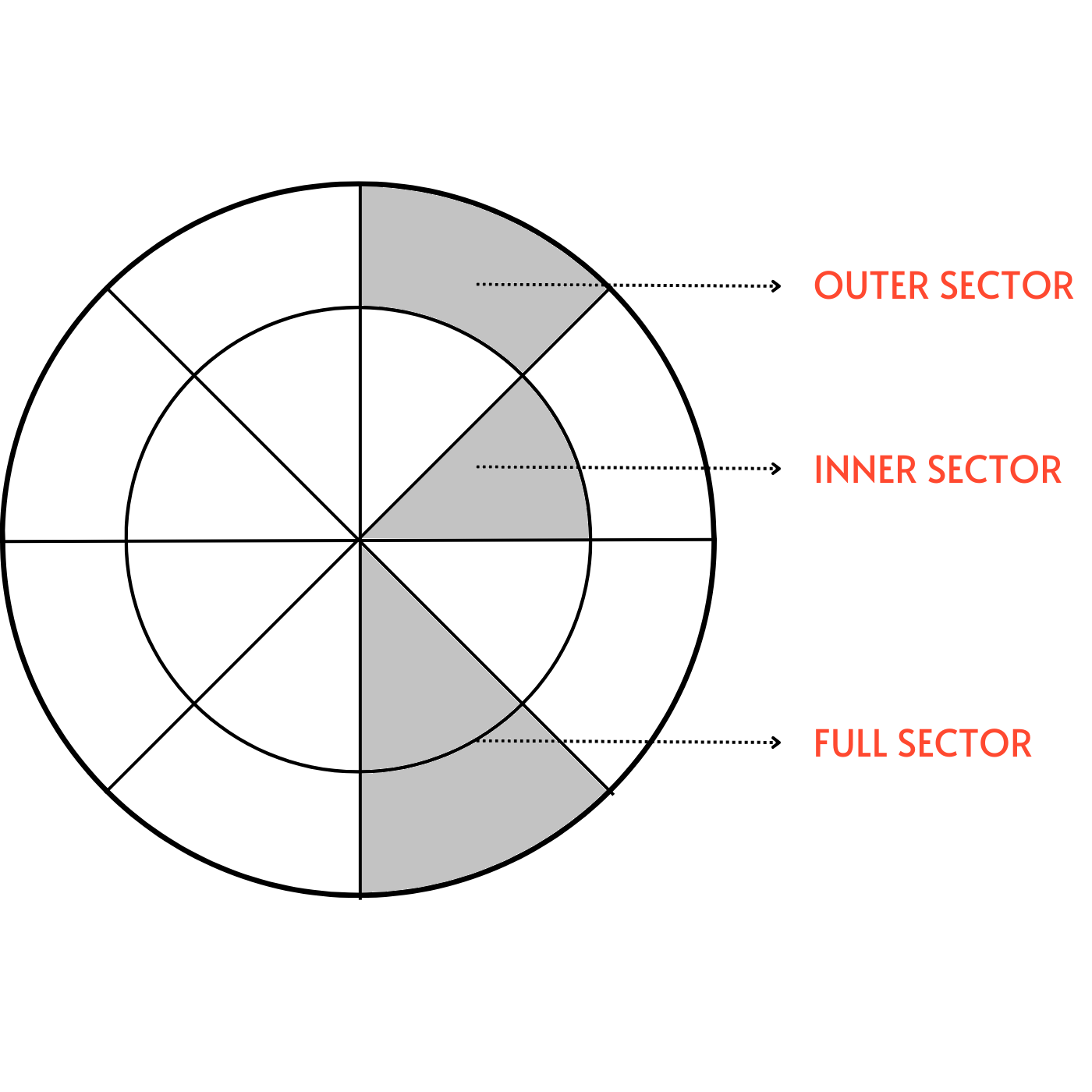}
  \caption{A schematic of the regions that could be involved in a sectorial trigger. The setup thus established can fit easily with the annular pinning scheme. However, to isolate the effect of triggers, we consider a uniform pinning strength throughout the star in the relevant simulations presented here.}
  \label{fig:sectorial_schematic}
\end{figure}

We implement a mechanism by which several pinning sites in a sector of the star are turned off a prescribed number of times throughout the run. The choice of the sector, the exact pinning sites that are switched off, and the times at which such a procedure is carried out are all random. The number of sectors that a star should be divided into, the probability with which a site should be switched off within any chosen sector, and the duration for which the site should remain inactive are specified as parameters. Further, to easily work in tandem with the annular scheme, functionality to switch off the pinning either throughout a sector (full sector), or only across a part of it (inner or outer sector) has also been included. See Fig.~\ref{fig:sectorial_schematic}. The parameter values used are listed in Table~\ref{tab:parameters}.

We run two sets of simulations with uniform pinning strength across the star:

1) 40 triggers, each lasting for one time period of the star ($T_{0}$), switching off one of the eight full sectors randomly.

2) 60 triggers, each lasting for one time period of the star ($T_{0}$), switching off one of the eight outer sectors randomly.

Animation A2 shows the evolution of the vortex array due to a sectorial trigger and the associated change in the rate of rotation. Several vortices in one of the outer sectors unpin and move in an almost azimuthal fashion. Soon, vortices from the interior are seen to move out radially. The glitch thus recorded was one of the largest in that run of the simulation. This progression of events is understood as follows. Before the trigger, the vortices are all pinned. Upon switching off the pinning in a sector with a probability of 0.5 for every site, some of the vortices present in the region simply follow the induced bulk superfluid motion, which is azimuthal in nature. This in itself does not change the superfluid's overall rate of rotation. However, a rearrangement of vortices causes a redistribution of stress locally, which in turn creates avalanches beginning with those vortices in nearby regions that are close to the threshold of unpinning.

\begin{figure}
  \centering
  \includegraphics[width=\linewidth]{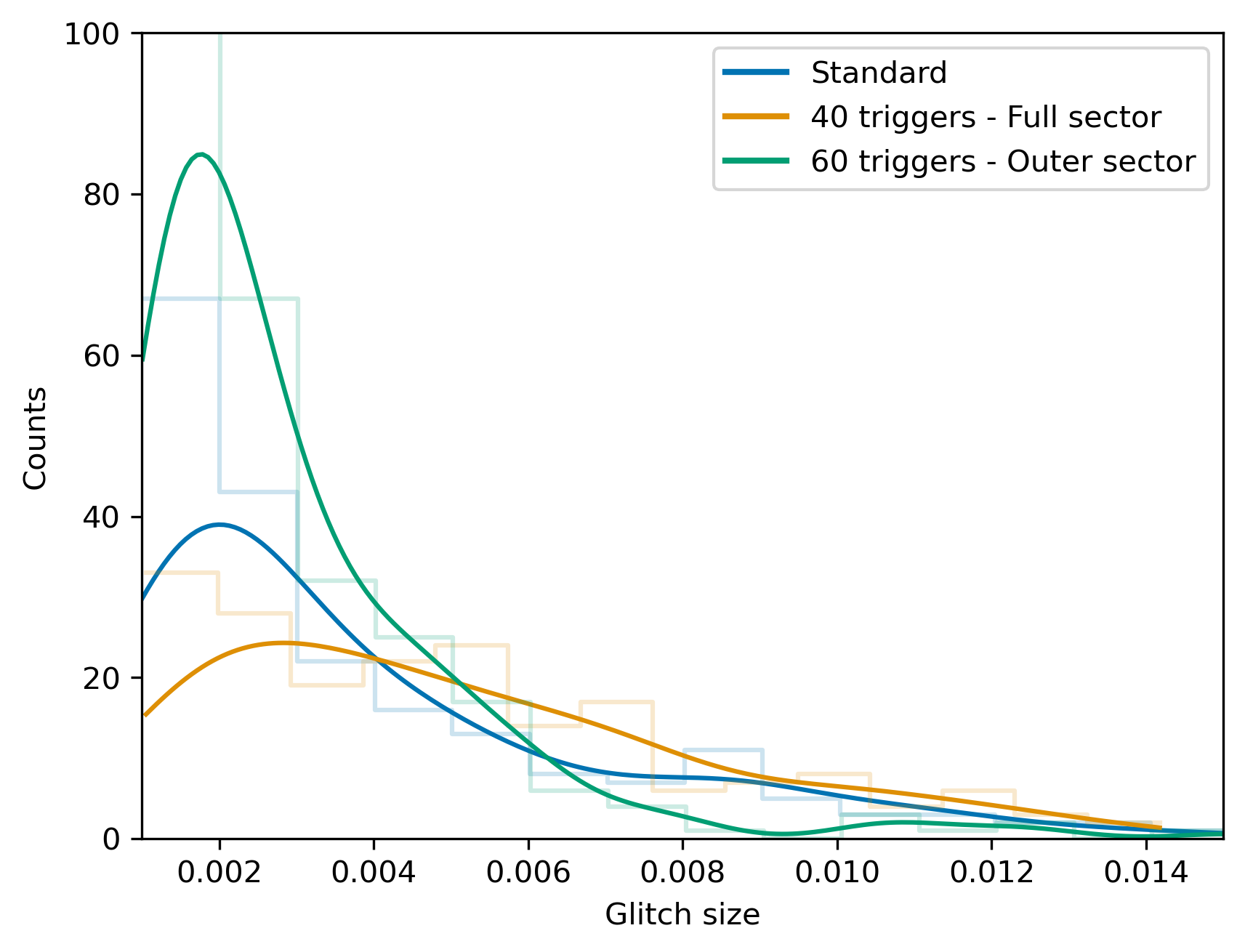}
  \caption{The distribution of large glitches resulting from simulations involving sectorial triggers.}
  \label{fig:sectorial_dist}
\end{figure}

Fig.~\ref{fig:sectorial_dist} shows that the inclusion of the trigger mechanism has visibly changed the distribution of the large glitches.

\subsection{Stress-waves}

Neutron stars can support both radial and nonradial modes of oscillation (\cite{mcdermott1988}). Such oscillations, if they leave their signatures on glitches, could provide crucial information about the insides of neutron stars (\cite{tran2023}). In isolated pulsars, oscillations may be initiated by crustquakes. In stars that are part of a binary system, accretion and nutation may also serve as causes (\cite{hoye1999}). But no matter the cause, such a normal mode of oscillation would increase the stress across the star in a periodic fashion, restricted in its duration due to damping mechanisms. This could lead to the unpinning of only the highly stressed vortices across the star. Once free to move, they would approach other vortices, increase the velocity field in the vicinity, and thus unpin them. Such a knock-on effect, along with the selective unpinning mentioned, holds the potential to create small avalanches across the star.

Such a phenomenon could be modelled in our setup by suitably altering the equation of motion. Instead, we take advantage of the module set up in the previous subsection to mimic the effect of stress-waves. We consider a trigger that unpins only those vortices with stresses greater than a set threshold ($\gamma$). This is achieved by decreasing the pinning strength across the star from $V_0 = 2000$ to $V_0 = \gamma * V_0$, for a short period of time ($T_{\mathrm{trig}}$). $\gamma$ is a factor with a value less than 1. Several of such triggers are placed randomly throughout the duration of the simulation. We run two sets of simulations with the parameter values detailed in Table~\ref{tab:parameters}. In one of these sets, during each trigger, the strength of the pinning sites is globally reduced to 0.7 times its original value. For the other set, the threshold value is 0.5.

Animation A3 shows the evolution of the vortex array due to a stress-wave and the associated change in the rate of rotation. The most stressed vortices are first released from the pinning sites once the trigger is activated. Some of them also knock out other pinned vortices, leading to further unpinning and migration.

\begin{figure}
  \centering
  \includegraphics[width=\linewidth]{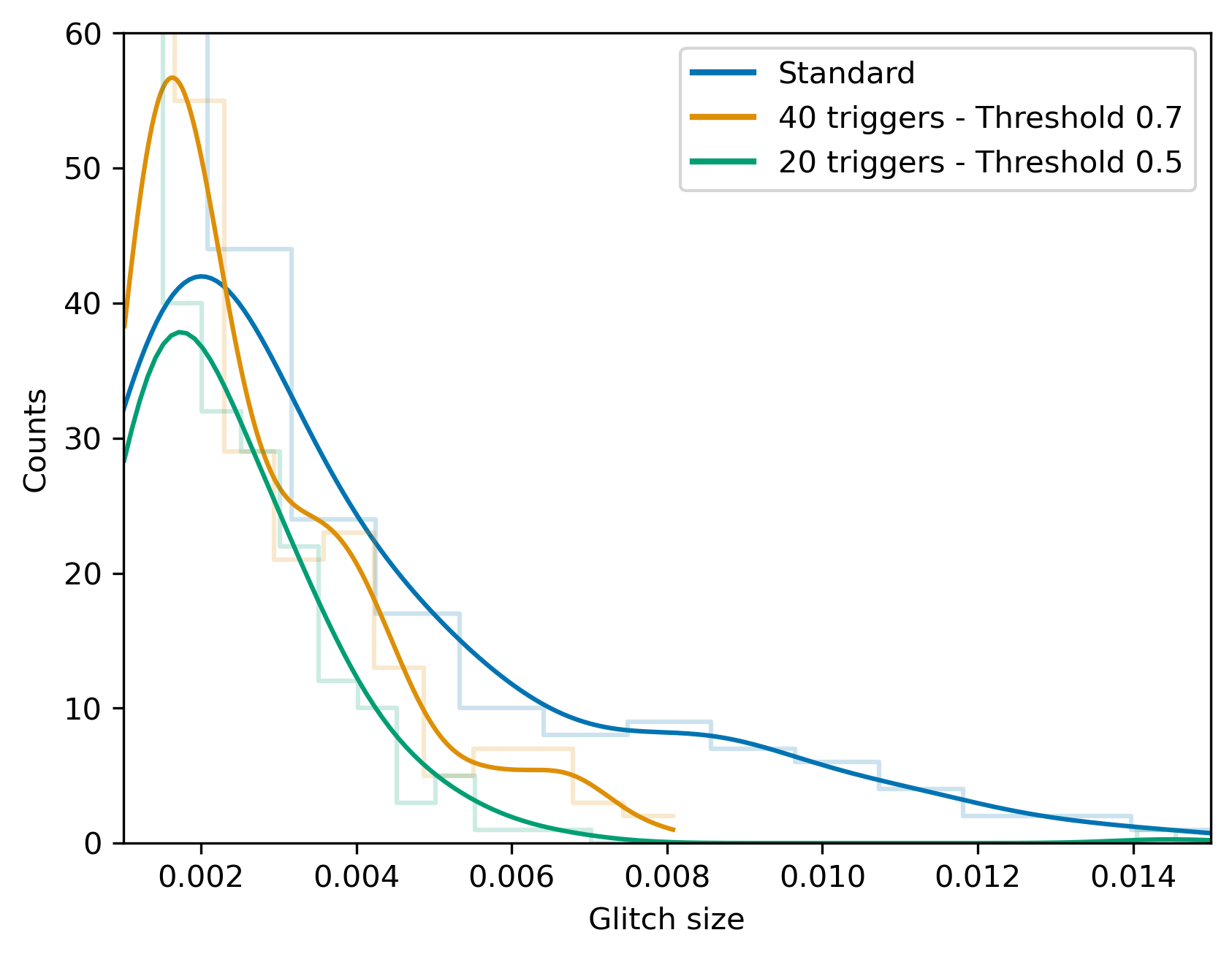}
  \caption{The distribution of large glitches resulting from simulations involving a stress-wave trigger.}
  \label{fig:stresswave_distributions}
\end{figure}

The resulting size distribution of large glitches presented in Fig.~\ref{fig:stresswave_distributions} carries signatures of the stress-wave.

\subsection{Hunting for bimodality}

The distributions presented in the preceding subsections possess clear evidence of the implemented variations. To identify those that stray well away from unimodality, we use Hartigan's Dip Test. The measures obtained thus are collected in Table~\ref{tab:hdp_simulations}.

\subsubsection{Annular pinning}

A strong bimodality with comparable peaks is observed in the annular pinning case with the outer annulus having stronger pinning than the inside ($V_1 = 1000 \text{~and~} V_{2}=2000$). But such a bimodality is not observed in the the scenario with $V_1 = 2000 \text{~and~} V_{2}=3000$. This may be attributed to the fact that setting $V_{2} = 3000$ leads to an increase in the maximum glitch size. This, alongside fewer big glitches, lends a wide distribution with an insignificant second peak. Thus, we may expect the two comparable peaks in the observed data of PSR J0537-6910 to place bounds on the radial distribution of pinning strengths.

\subsubsection{Sectorial triggers}

A weak bimodality is present in one of the sectorial trigger simulations (40 triggers - Full sector). At present, the effect of the trigger region and the number of triggers on the distribution is not well understood. Further insights into such connections will help narrow the subset of triggers that may produce a bimodal glitch-size distribution.

\subsubsection{Stress-waves}

No bimodality is recorded in the case of stress-waves. In these simulations, knock-on occurs only in a few locations in the course of the animated glitch. The major contribution to the glitch comes from the initially unpinned vortices. This behaviour is expected to be significantly different for simulations with a greater number density of vortices, where the knock-on processes could lead to several small avalanches progressing independently of each other. Such glitches would comprise several independent log-normal processes whose added effects would contribute to the size of the glitch. The Central Limit Theorem, then, implies a Gaussian distribution for the associated glitch-size. This would lend a lognormal-Gaussian mixture when both stress-wave glitches and Magnus-induced glitches are present in the star. A confirmation of this using the current setup is computationally expensive but within reach, making it suitable for exploration in a future communication.

\begin{table}
\centering
\caption{Hartigan's dip statistics corresponding to the distributions of large glitches resulting from various simulations. Underlined entries indicate a possible departure from unimodality.}
\label{tab:hdp_simulations}
\begin{tabular}{@{}lll@{}}
\toprule
\textbf{Variation} & \textbf{Dip statistic} & \textbf{p-value}\\ \midrule
Standard             & 0.018                 & 0.950\\
\midrule
Annular pinning             &                       &      \\
1000, 2000           & \ul{0.056}                 & \ul{0.001}\\
2000, 1000           & 0.014                 & 0.994\\
3000, 2000           & 0.018                 & 0.958\\
2000, 3000           & 0.014                 & 0.999\\
\midrule
Sectorial triggers           &                       &      \\
40 triggers - Full sector           & \ul{0.024}                 & \ul{0.628}\\ 
60 triggers - Outer sector           & 0.013                 & 0.993\\ 
\midrule
Stress-waves            &                       &      \\
40 triggers - Threshold 0.7           & 0.016                 & 0.974\\ 
20 triggers - Threshold 0.5           & 0.017                 & 0.968\\ 
\bottomrule
\end{tabular}
\end{table}

\subsection{Remarks}

The essential results have been reported in the previous subsections. However, two features not mentioned above are worth drawing attention to.

\subsubsection{Smoothing}

\begin{figure}
  \centering
    \includegraphics[width=\linewidth]{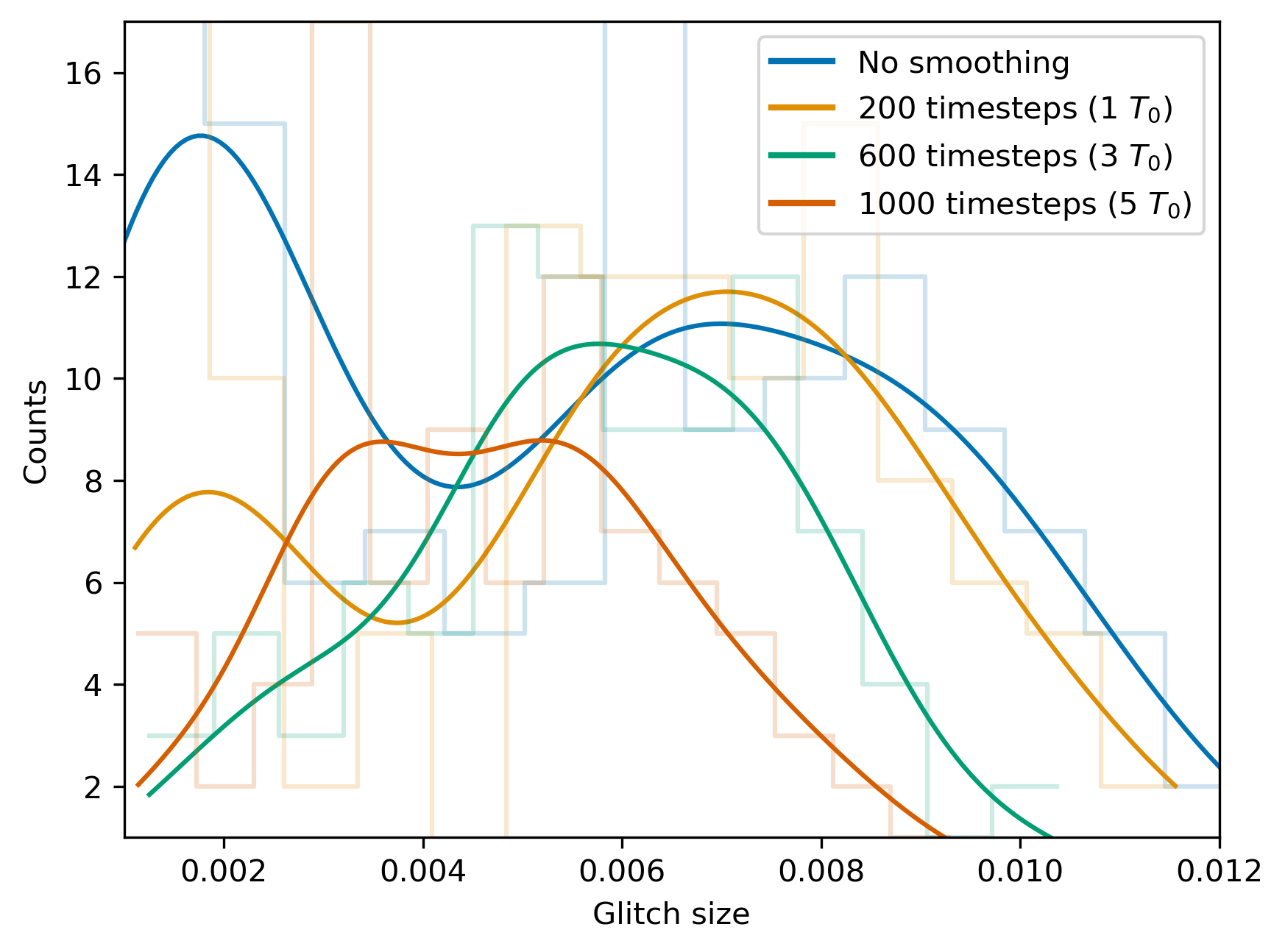}
  \caption{The effect of observational smoothing on an originally bimodal glitch-size distribution.}
  \label{fig:smoothened_annular_1000_2000}
\end{figure}

The rotation rate of the star is read from the simulator at each time step. Pulsars, on the other hand, are not as regularly monitored. This leads to a smoothing of the observations, thus biasing the data against small glitches that have a fast rise and decay. In order to emulate the effect of such smoothing on an originally bimodal distribution, the rotation rate data corresponding to the annular pinning simulation with $V_1 = 1000$ and $V_2 = 2000$ has been smoothed over 200 timesteps ($1 ~ T_{0}$), 600 timesteps ($3 ~  T_{0}$), and 1000 timesteps ($5 ~ T_{0}$), respectively. The resulting distributions are presented in Fig.~\ref{fig:smoothened_annular_1000_2000}. The bimodality declines in strength when the smoothing factor is very high. This suggests that multimodality in glitch-size distributions of several pulsars could  be suppressed due to infrequent observations.

\subsubsection{Rise of glitches}

\begin{figure}
  \centering
  \includegraphics[width=\linewidth]{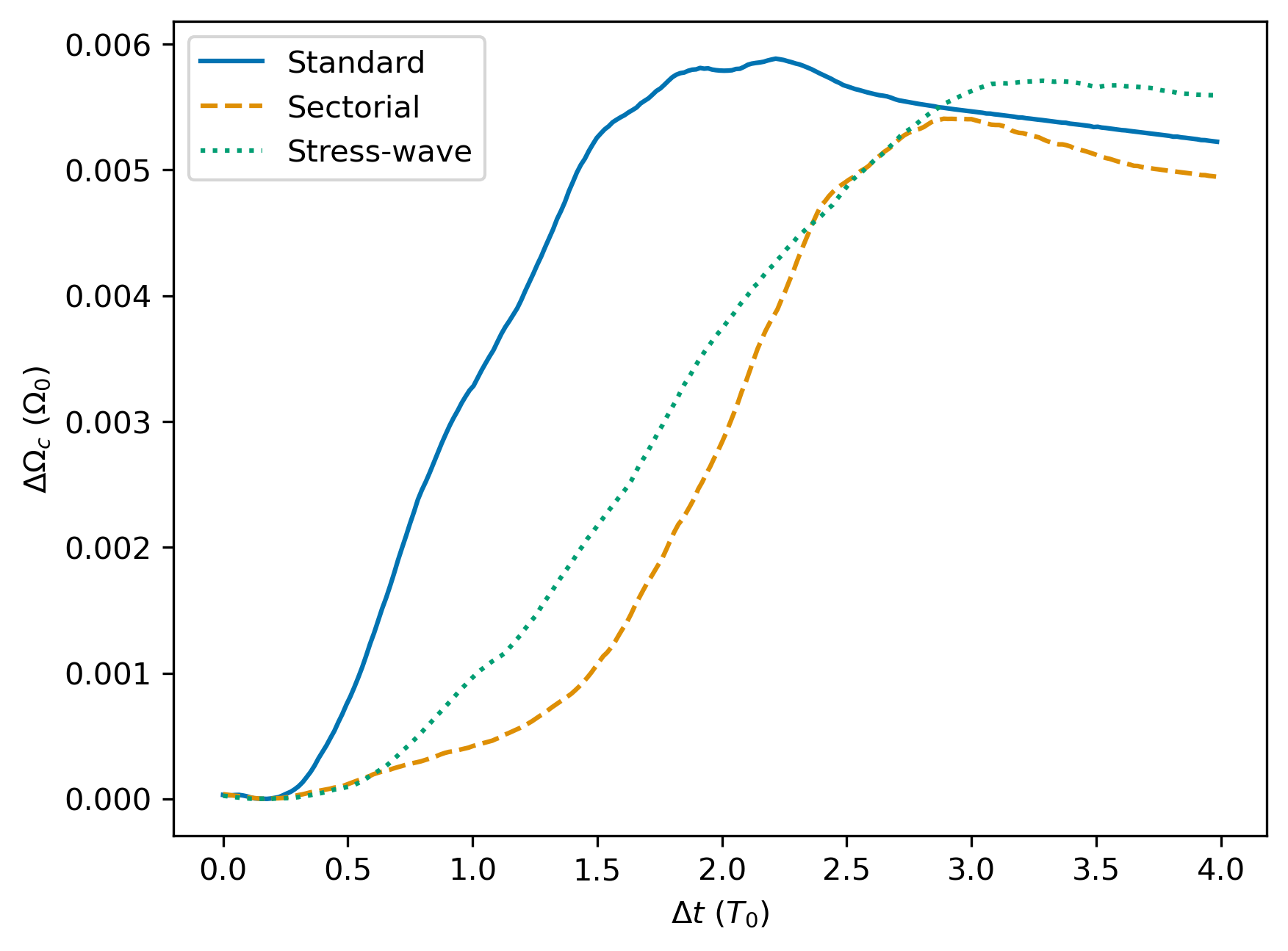}
  \caption{Comparison of the rise of glitches having similar sizes, occurring in the standard, sectorial, and stress-wave simulations, respectively, at late times.}
  \label{fig:glitchrise}
\end{figure}

The qualitative difference in the glitch-rise between the various models is notable. This is illustrated through the animations. In the standard case, the rise is rather gradual. However, in the triggered cases, the rise is sharper. Upon checking several glitches of similar sizes, we find that this behaviour cannot be generalised. We see that the triggered glitches, in fact, have a slower rise if they occur late in the simulation, after stresses have been repeatedly released by the previous glitches. See Fig.~\ref{fig:glitchrise}. The approach presented in this section, thus, holds promise in studying the varied rise and recovery characteristics of individual pulsars, an active question of research concerning the physics of neutron stars.

\section{Conclusions}

The vast variety of features exhibited by pulsar glitches and their associated statistics have been difficult to reconcile within a single model. We focus on the pulsar with the most observed glitches, PSR J0537-6910, and analyse its glitch-size distribution. We find a bimodality. Using recently proposed computational techniques (\cite{howitt2020}), we simulate a vortex array and include effects that mimic scenarios relevant to a neutron star. We deduce that annular variations in pinning can produce an evident bimodality in the observed distribution. Processes which occasionally unpin several vortices in a localised region could produce a weak bimodality. Global disturbances do not result in any such features for the range of parameters considered. However, strong signatures of the variations are present. Thus, this preliminary study stitches together structural inhomogeneities, trigger mechanisms, superfluid vortex arrays, and glitch statistics into a unified, albeit incomplete, tapestry requiring further attention. In addition, we also suggest that the observed variation in the rise of glitches possibly reflects the underlying trigger, an idea that may be probed further using the computational methods demonstrated above.

At present, it may be possible to study only a handful of pulsars reliably owing to the sparse data available. But prudent categorisation of pulsars to obtain a larger dataset may somewhat alleviate the situation. In addition, analysis performed with a small sample size is susceptible to drastic changes with the inclusion of new observations. Hence, new techniques are required to study the stability of slowly growing distributions.

Furthermore, the core of the neutron star, with three-dimensional phenomena such as tangling of vortices and flux tubes, may not be well described by two-dimensional models. The effect of the vortex array on the flux tubes and the converse are both not well understood. The strong magnetic field of a neutron star is also expected to have a considerable impact on the crustal properties. The relative importance of such considerations in the context of glitch statistics merits future investigation.

\section*{Acknowledgements}

The authors thank Ashoka University for continued support. SVA thanks Amin Nizami, Debarati Chatterjee, Kandaswamy Subramanian, Gautam Menon, George Howitt, Brynmor Haskell, and his peers for helpful discussions and comments at various stages of the work. The simulations in this paper were run on the Chanakya HPC system at Ashoka University. The animations were prepared using the Manim package available for python (\cite{manim2024}). The HDT statistics were calculated on python using the diptest package.

\section*{Data availability}

The observational data used in this study can be accessed from \href{https://www.jb.man.ac.uk/pulsar/glitches/gTable.html}{Jodrell Bank Glitch Catalogue}.
The simulation data used in this article can be made available upon request by emailing the corresponding author. However, the codes developed by the authors have been made publicly available through the following github repository: \href{https://github.com/ananth-94/vortexsimulator.git}{Vortex simulator SVA-DB}. They can be used by appropriately citing this article.



\bibliographystyle{mnras}
\bibliography{references} 

\begin{thebibliography}{}
\makeatletter
\relax
\def\mn@urlcharsother{\let\do\@makeother \do\$\do\&\do\#\do\^\do\_\do\%\do\~}
\def\mn@doi{\begingroup\mn@urlcharsother \@ifnextchar [ {\mn@doi@} {\mn@doi@[]}}
\def\mn@doi@[#1]#2{\def\@tempa{#1}\ifx\@tempa\@empty \href {http://dx.doi.org/#2} {doi:#2}\else \href {http://dx.doi.org/#2} {#1}\fi \endgroup}
\def\mn@eprint#1#2{\mn@eprint@#1:#2::\@nil}
\def\mn@eprint@arXiv#1{\href {http://arxiv.org/abs/#1} {{\tt arXiv:#1}}}
\def\mn@eprint@dblp#1{\href {http://dblp.uni-trier.de/rec/bibtex/#1.xml} {dblp:#1}}
\def\mn@eprint@#1:#2:#3:#4\@nil{\def\@tempa {#1}\def\@tempb {#2}\def\@tempc {#3}\ifx \@tempc \@empty \let \@tempc \@tempb \let \@tempb \@tempa \fi \ifx \@tempb \@empty \def\@tempb {arXiv}\fi \@ifundefined {mn@eprint@\@tempb}{\@tempb:\@tempc}{\expandafter \expandafter \csname mn@eprint@\@tempb\endcsname \expandafter{\@tempc}}}

\bibitem[\protect\citeauthoryear{Andersson, Glampedakis, Ho  \& Espinoza}{Andersson et~al.}{2012}]{andersson2012}
Andersson N.,  Glampedakis K.,  Ho W. C.~G.,   Espinoza C.~M.,  2012, \mn@doi [Physical Review Letters] {10.1103/PhysRevLett.109.241103}, 109, 241103

\bibitem[\protect\citeauthoryear{Antonopoulou, Haskell  \& Espinoza}{Antonopoulou et~al.}{2022}]{antonopoulou2022}
Antonopoulou D.,  Haskell B.,   Espinoza C.~M.,  2022, \mn@doi [Reports on Progress in Physics] {10.1088/1361-6633/ac9ced}, 85, 126901

\bibitem[\protect\citeauthoryear{Bak, Tang  \& Wiesenfeld}{Bak et~al.}{1987}]{bak1987}
Bak P.,  Tang C.,   Wiesenfeld K.,  1987, \mn@doi [Physical Review Letters] {10.1103/PhysRevLett.59.381}, 59, 381

\bibitem[\protect\citeauthoryear{Basu et~al.,}{Basu et~al.}{2022}]{basu2022}
Basu A.,  et~al., 2022, \mn@doi [Monthly Notices of the Royal Astronomical Society] {10.1093/mnras/stab3336}, 510, 4049

\bibitem[\protect\citeauthoryear{Ceva \& Luzuriaga}{Ceva \& Luzuriaga}{1998}]{ceva1998}
Ceva H.,  Luzuriaga J.,  1998, \mn@doi [Physics Letters A] {10.1016/S0375-9601(98)00848-2}, 250, 275

\bibitem[\protect\citeauthoryear{Espinoza, Antonopoulou, Stappers, Watts  \& Lyne}{Espinoza et~al.}{2014}]{espinoza2014}
Espinoza C.~M.,  Antonopoulou D.,  Stappers B.~W.,  Watts A.,   Lyne A.~G.,  2014, \mn@doi [Monthly Notices of the Royal Astronomical Society] {10.1093/mnras/stu395}, 440, 2755

\bibitem[\protect\citeauthoryear{Eya, Urama  \& Chukwude}{Eya et~al.}{2019}]{eya2019}
Eya I.~O.,  Urama J.~O.,   Chukwude A.~E.,  2019, \mn@doi [Research in Astronomy and Astrophysics] {10.1088/1674-4527/19/6/89}, 19, 089

\bibitem[\protect\citeauthoryear{Fuentes, Espinoza, Reisenegger, Stappers, Shaw  \& Lyne}{Fuentes et~al.}{2017}]{fuentes2017}
Fuentes J.~R.,  Espinoza C.~M.,  Reisenegger A.,  Stappers B.~W.,  Shaw B.,   Lyne A.~G.,  2017, \mn@doi [Astronomy \& Astrophysics] {10.1051/0004-6361/201731519}, 608, A131

\bibitem[\protect\citeauthoryear{Fuentes, Espinoza  \& Reisenegger}{Fuentes et~al.}{2019}]{fuentes2019}
Fuentes J.~R.,  Espinoza C.~M.,   Reisenegger A.,  2019, \mn@doi [Astronomy \& Astrophysics] {10.1051/0004-6361/201935939}, 630, A115

\bibitem[\protect\citeauthoryear{Glampedakis \& Andersson}{Glampedakis \& Andersson}{2009}]{glampedakis2008}
Glampedakis K.,  Andersson N.,  2009, \mn@doi [Physical Review Letters] {10.1103/PhysRevLett.102.141101}, 102, 141101

\bibitem[\protect\citeauthoryear{Hartigan}{Hartigan}{1985}]{hartigan1985_2}
Hartigan P.,  1985, Journal of the Royal Statistical Society. Series C (Applied Statistics), 34, 320

\bibitem[\protect\citeauthoryear{Hartigan \& Hartigan}{Hartigan \& Hartigan}{1985}]{hartigan1985_1}
Hartigan J.~A.,  Hartigan P.~M.,  1985, The annals of Statistics, pp 70--84

\bibitem[\protect\citeauthoryear{Haskell \& Melatos}{Haskell \& Melatos}{2015}]{haskell2015}
Haskell B.,  Melatos A.,  2015, \mn@doi [International Journal of Modern Physics D] {10.1142/S0218271815300086}, 24, 1530008

\bibitem[\protect\citeauthoryear{Hays \& Winkler}{Hays \& Winkler}{1971}]{hays1971}
Hays W.~L.,  Winkler R.~L.,  1971, Statistics: Probability, Inference, and Decision.
Series in Quantitative Methods for Decision-Making, {Holt, Rinehart and Winston}, New York, NY

\bibitem[\protect\citeauthoryear{Howitt, Melatos  \& Delaigle}{Howitt et~al.}{2018}]{howitt2018}
Howitt G.,  Melatos A.,   Delaigle A.,  2018, \mn@doi [The Astrophysical Journal] {10.3847/1538-4357/aae20a}, 867, 60

\bibitem[\protect\citeauthoryear{Howitt, Melatos  \& Haskell}{Howitt et~al.}{2020}]{howitt2020}
Howitt G.,  Melatos A.,   Haskell B.,  2020, \mn@doi [Monthly Notices of the Royal Astronomical Society] {10.1093/mnras/staa2314}, 498, 320

\bibitem[\protect\citeauthoryear{Hoye}{Hoye}{1999}]{hoye1999}
Hoye G.~K.,  1999, PhD thesis, Norwegian University of Science and Technology, Trondheim

\bibitem[\protect\citeauthoryear{Kass \& Raftery}{Kass \& Raftery}{1995}]{kass1995}
Kass R.~E.,  Raftery A.~E.,  1995, \mn@doi [Journal of the American Statistical Association] {10.1080/01621459.1995.10476572}, 90, 773

\bibitem[\protect\citeauthoryear{Khomenko, Antonelli  \& Haskell}{Khomenko et~al.}{2019}]{khomenko2019}
Khomenko V.,  Antonelli M.,   Haskell B.,  2019, \mn@doi [Physical Review D] {10.1103/PhysRevD.100.123002}, 100, 123002

\bibitem[\protect\citeauthoryear{Konar \& Arjunwadkar}{Konar \& Arjunwadkar}{2014}]{konar2014}
Konar S.,  Arjunwadkar M.,  2014, Bull. Astr. Soc. India

\bibitem[\protect\citeauthoryear{Liu, Baggaley, Barenghi  \& Wood}{Liu et~al.}{2024}]{liu2024}
Liu I.-K.,  Baggaley A.~W.,  Barenghi C.~F.,   Wood T.~S.,  2024, Vortex Avalanches and Collective Motion in Neutron Stars (\mn@eprint {arXiv} {2410.16878}), \url {https://arxiv.org/abs/2410.16878}

\bibitem[\protect\citeauthoryear{McDermott, {van Horn}  \& Hansen}{McDermott et~al.}{1988}]{mcdermott1988}
McDermott P.~N.,  {van Horn} H.~M.,   Hansen C.~J.,  1988, \mn@doi [The Astrophysical Journal] {10.1086/166044}, 325, 725

\bibitem[\protect\citeauthoryear{Mitzenmacher}{Mitzenmacher}{2004}]{mitzenmacher2004}
Mitzenmacher M.,  2004, \mn@doi [Internet Mathematics] {10.1080/15427951.2004.10129088}, 1, 226

\bibitem[\protect\citeauthoryear{Montroll \& Shlesinger}{Montroll \& Shlesinger}{1982}]{montroll1982}
Montroll E.~W.,  Shlesinger M.~F.,  1982, Proceedings of the National Academy of Sciences of the United States of America, 79, 3380

\bibitem[\protect\citeauthoryear{Newville, Stensitzki, Allen  \& Ingargiola}{Newville et~al.}{2015}]{newville2015}
Newville M.,  Stensitzki T.,  Allen D.~B.,   Ingargiola A.,  2015, {LMFIT: Non-Linear Least-Square Minimization and Curve-Fitting for Python}, \mn@doi{10.5281/zenodo.11813}, \url {https://doi.org/10.5281/zenodo.11813}

\bibitem[\protect\citeauthoryear{Parzen}{Parzen}{1962}]{parzen1962}
Parzen E.,  1962, \mn@doi [The Annals of Mathematical Statistics] {10.1214/aoms/1177704472}, 33, 1065

\bibitem[\protect\citeauthoryear{Peralta, Melatos, Giacobello  \& Ooi}{Peralta et~al.}{2006}]{peralta2006}
Peralta C.,  Melatos A.,  Giacobello M.,   Ooi A.,  2006, \mn@doi [The Astrophysical Journal] {10.1086/507576}, 651, 1079

\bibitem[\protect\citeauthoryear{Reed \& Jorgensen}{Reed \& Jorgensen}{2004}]{reed2004}
Reed W.~J.,  Jorgensen M.,  2004, \mn@doi [Communications in Statistics - Theory and Methods] {10.1081/STA-120037438}, 33, 1733

\bibitem[\protect\citeauthoryear{Rencoret, {Aguilera-G{\'o}mez}  \& Reisenegger}{Rencoret et~al.}{2021}]{rencoret2021}
Rencoret J.~A.,  {Aguilera-G{\'o}mez} C.,   Reisenegger A.,  2021, \mn@doi [Astronomy \& Astrophysics] {10.1051/0004-6361/202141499}, 654, A47

\bibitem[\protect\citeauthoryear{Ruderman}{Ruderman}{1969}]{ruderman1969}
Ruderman M.,  1969, \mn@doi [Nature] {10.1038/223597b0}, 223, 597

\bibitem[\protect\citeauthoryear{Sauls}{Sauls}{1989}]{sauls1989}
Sauls J.~A.,  1989, in {\"O}gelman H.,  {van den Heuvel} E. P.~J.,  eds, , Timing {{Neutron Stars}}.
Springer Netherlands, Dordrecht, pp 457--490, \mn@doi{10.1007/978-94-009-2273-0_43}

\bibitem[\protect\citeauthoryear{Schwarz}{Schwarz}{1978}]{schwarz1978}
Schwarz G.,  1978, \mn@doi [The Annals of Statistics] {10.1214/aos/1176344136}, 6, 461

\bibitem[\protect\citeauthoryear{Srinivasan, Bhattacharya, Muslimov  \& Tsygan}{Srinivasan et~al.}{1990}]{srinivasan1990}
Srinivasan G.,  Bhattacharya D.,  Muslimov A.~G.,   Tsygan A.~I.,  1990, Current Science, 59, 31

\bibitem[\protect\citeauthoryear{{The Manim Community Developers}}{{The Manim Community Developers}}{2024}]{manim2024}
{The Manim Community Developers} 2024, {Manim – Mathematical Animation Framework}, \url {https://www.manim.community/}

\bibitem[\protect\citeauthoryear{Tran, Ghosh, Lozano, Chatterjee  \& Jaikumar}{Tran et~al.}{2023}]{tran2023}
Tran V.,  Ghosh S.,  Lozano N.,  Chatterjee D.,   Jaikumar P.,  2023, \mn@doi [Physical Review C] {10.1103/PhysRevC.108.015803}, 108, 015803

\bibitem[\protect\citeauthoryear{Warszawski \& Melatos}{Warszawski \& Melatos}{2008}]{warszawski2008}
Warszawski L.,  Melatos A.,  2008, \mn@doi [Monthly Notices of the Royal Astronomical Society] {10.1111/j.1365-2966.2008.13662.x}, 390, 175

\bibitem[\protect\citeauthoryear{Warszawski \& Melatos}{Warszawski \& Melatos}{2011}]{warszawski2011}
Warszawski L.,  Melatos A.,  2011, \mn@doi [Monthly Notices of the Royal Astronomical Society] {10.1111/j.1365-2966.2011.18803.x}, 415, 1611

\makeatother
\end{thebibliography}



\appendix

\section{Theoretical reconciliation}

An exact quantitative connection between vortex dynamics and the resulting glitch-size distributions has not yet been established. However, for the model presented above, \cite{howitt2020} note that no strong cross-correlations  were found between the glitch sizes and the waiting times, suggesting the presence of self-organised criticality (SOC), counterintuitive to the threshold-triggered stress-release dynamics integrated within. With this as our guide, in the following discussion, we identify general principles defining the model and extract statistical features from the same.

If we consider the evolution of the vortex array as a random stress addition and redistribution process, we can draw parallels between the array and a sand-pile as defined in \cite{bak1987}. The neutron star as a sand-pile is an idea that has been previously explored in \cite{warszawski2008}, where the size distribution of the glitches is shown to follow a power law. In the following, we briefly explain the sand-pile model and elucidate its relevance to understanding the distributions resulting from our simulations.

The sand-pile model is a computational paradigm in which grains are added to random points on a finite square lattice. There exists a threshold beyond which a stack tumbles onto its neighbouring sites. After several of such redistributions, the system reaches a critical state where the addition of one grain could lead to a series of topples across the system. The size distribution of such avalanches in this self-organized critical state is found to be a power law.

A crucial and not widely quoted observation on the sand-pile model is presented in \cite{ceva1998}: Avalanches (of a fixed duration) occurring early in the sand-pile simulations follow a log-normal distribution, due to the probabilities involved being independent of each other. Over time, after many such avalanches have occurred, correlations are established in the system, and a power-law distribution is observed.

\begin{figure}
  \centering
  \includegraphics[width=\linewidth]{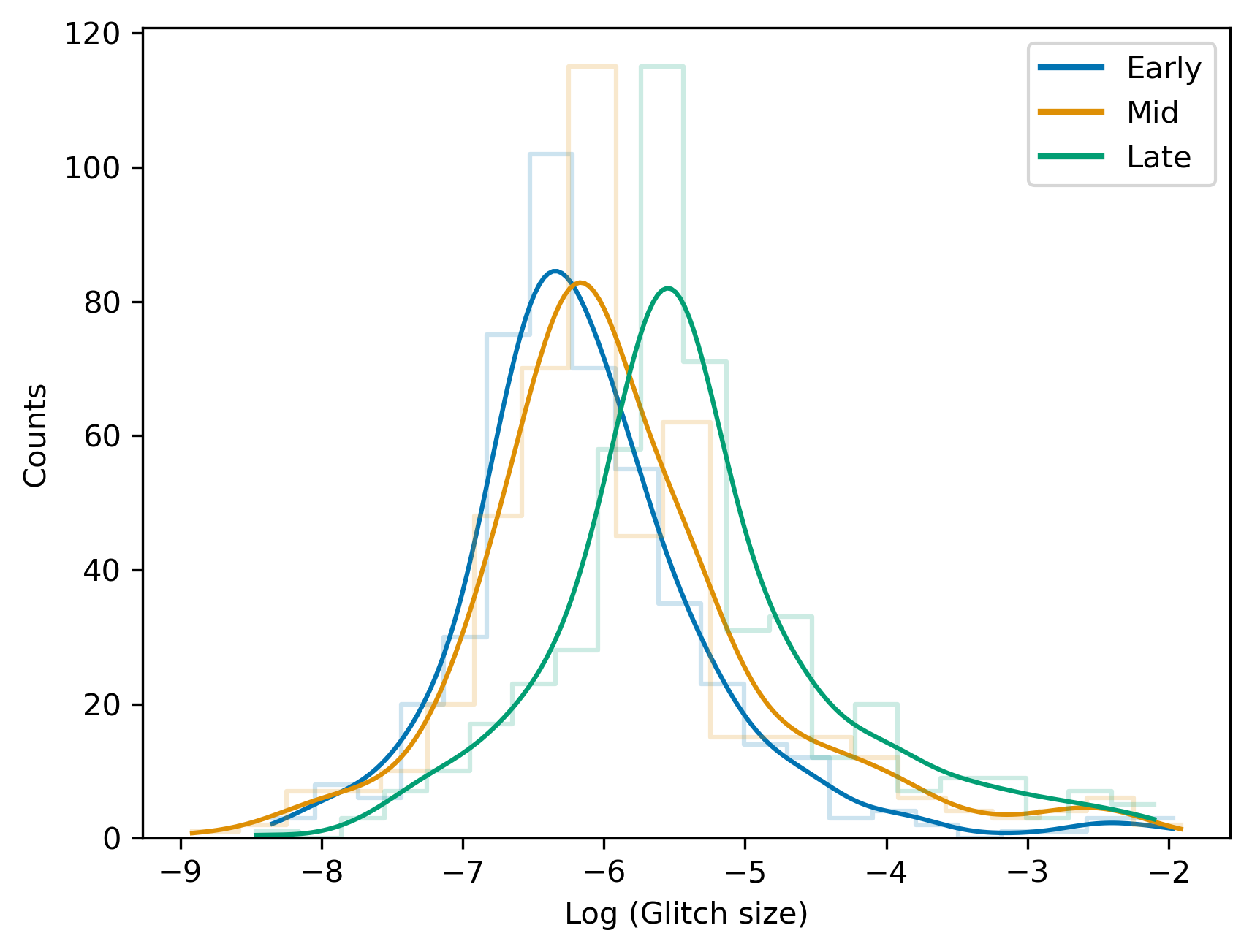}
  \caption{The distribution of glitches at early, mid, and late times of the standard simulation. Early, mid and late times are defined as being small windows of time around $0.1~T_{\mathrm{run}}$, $0.5~ T_{\mathrm{run}}$, and $0.9~T_{\mathrm{run}}$, respectively.}
  \label{fig:sandpile}
\end{figure}

We compare the early, mid, and late time distributions resulting from the standard simulations in Fig.~\ref{fig:sandpile}. The distributions are all similar, except for the mode. No transition from log-normal to power-law is observed. This suggests that the model of the neutron star used is either not a system that can be modelled as a sand-pile or corresponds to a system that is in the log-normal (uncorrelated) regime throughout the simulations.

The latter possibility cannot be dismissed easily, since the distributions corresponding to the standard simulations have been shown to follow the log-normal. To understand this, we note that there exists a subtle difference between a sand-pile and a vortex array. We identify grains in a sand-pile with stresses in the neutron star. In the sand-pile  model, in the bulk, the number of grains remains conserved. Thus, only a simple redistribution occurs. On the other hand, in the model of the star used, the stresses are not conserved. During an avalanche, the stresses are redistributed in a short period of time. But a glitch causes the Magnus force, and thus the stresses, to reduce throughout the star, effectively resetting the initial conditions and possibly destroying the onset of a correlation. Thus, every glitch in the standard case would be derived from a log-normal distribution of size, corresponding to the length of the event.

In our simulations, as in a pulsar, the durations of the avalanches vary. The size of every avalanche would then be drawn from a slightly different log-normal distribution. If the durations themselves are random and are drawn from an exponential distribution, then the associated size distribution has been shown to follow a double Pareto-lognormal, where the body of the distribution is fit better by a log-normal and the tail by a power-law (\cite{montroll1982, reed2004, mitzenmacher2004}). This behaviour has been observed in our simulations, as discussed in Section 3.2.

Thus, we find that generative models of distributions, when combined prudently with general physical principles, can offer promising insights connecting the collective microscopic behaviour of a system to macroscopic effects.

\bsp	
\label{lastpage}
\end{document}